\documentclass[aps,prl,twocolumn,showpacs,superscriptaddress,groupedaddress]{revtex4}  

\usepackage{amssymb}
\usepackage{amsmath}
\usepackage{epsfig}
\usepackage{epstopdf}
\usepackage{bm}
\usepackage{graphicx,epsfig}
\usepackage{mathrsfs}
\usepackage{dcolumn}
\usepackage{color}
\usepackage{natbib}
\usepackage{CJK}
\usepackage{tikz}
\usetikzlibrary{arrows,shapes,trees}
\def\be{\begin{equation}}
\def\ee{\end{equation}}
\def\bea{\begin{eqnarray}}
\def\eea{\end{eqnarray}}
\def\la{\langle}
\def\ra{\rangle}

\allowdisplaybreaks[2]

\begin{document}

\title{Heisenberg Limit Superradiant Super-resolving Metrology}
\author{Da-Wei Wang}
\affiliation{Texas A$\&$M University, College Station, TX 77843, USA}
\author{Marlan O. Scully}
\affiliation{Texas A$\&$M University, College Station, TX 77843, USA}
\affiliation{Princeton University, Princeton, New Jersey 08544, USA}
\affiliation{Baylor University, Waco, TX 76706, USA}

\date{\today }

\begin{abstract}
We propose a superradiant metrology technique to achieve the Heisenberg limit super-resolving displacement measurement by encoding multiple light momenta into a three-level atomic ensemble. We use $2N$ coherent pulses to prepare a single excitation superradiant state in a superposition of two timed Dicke states that are $4N$ light momenta apart in momentum space. The phase difference between these two states induced by a uniform displacement of the atomic ensemble has $1/4N$ sensitivity. Experiments are proposed in crystals and in ultracold atoms.
\end{abstract}


\maketitle

\emph{Introduction}.---Measurements play a key role in physics, not only in the direct sense of defining standards \cite{Kasevich1989, Huelga1997, Borregaard2013}, but also in verifying predictions of theories such as gravitational waves \cite{Abramovici1992, Thorne1980}. Using $N$ quantum resources independently, the sensitivity of a parameter $\phi$ scales $\Delta\phi\sim 1/\sqrt{N}$ determined by the central limit theorem. This $1/\sqrt{N}$ scaling is the so-called shot-noise limit \cite{Xiao1987, Giovannetti2011}, which can be broken by squeezed states or entanglement \cite{Caves1981, Yurke1986, Giovannetti2004, Giovannetti2006, Pezze2008, Xiang2011, AasiJ2013}. For example, the photons can be prepared in a Schr\"odinger cat state, such as the N00N state \cite{Boto2000}, 
$1/\sqrt{2}\left(|N0\ra+|0N\ra\right)$,
where $|N0\ra$ means all the photons are in one arm of the interferometer and $|0N\ra$ means all the photons are in the other arm. According to the unbiased Cram\'er-Rao bound, the sensitivity can be enhanced to $\Delta \phi\sim 1/N$, the Heisenberg limit. Another scheme to reach to the Heisenberg limit is to use the $N$-atom Schr\"odinger cat state, the so-called Greenberger-Horne-Zeilinger (GHZ) state \cite{Greenberger1990, Bollinger1996, Molmer1999, Leibfried2005, Jones2009}.

The metrology with N00N states and GHZ states indicates that we can improve the sensitivity of one observable by preparing a Schr\"odinger cat state of its conjugate observable \cite{Giovannetti2006}. The sensitivity of the displacement $x$ can therefore be improved by preparing a Schr\"odinger cat state of the momentum $p$. The N00N state composed by two optical modes with opposite momenta seems to be a good candidate, but the fragile high photon number N00N state is very difficult to prepare, to preserve, and to manipulate \cite{Mitchell2004, Walther2004, Nagata2007, Afek2010, Israel2012, McCusker2009, Rosen2012}. On the other hand, a single photon can be easily prepared in an entangled state of two opposite momenta, $1/{\sqrt{2}}\left(|1_\mathbf{k}0_\mathbf{-k}\ra+|0_\mathbf{k}1_\mathbf{-k}\ra\right)$ \cite{Higgins2007}. To manipulate this entangled state, we can guide it to a collection of atoms and the momentum can be translated into the phase of the collective excitation, $1/{\sqrt{2}}\left(|b_\mathbf{k}\ra+|b_\mathbf{-k}\ra\right)$, 
where 
\begin{equation}
\left|b_{\mathbf{k}}\right>\equiv\frac{1}{\sqrt{N_a}}\sum_{j=1}^{N_a}\exp\left(ik x_j\right)\left|c_1,c_2,...,b_j,...c_{N_a}\right>
\label{sr}
\end{equation} 
is the timed Dicke state \cite{Scully2006} and $N_a$ is the number of atoms. Here we prepare a pencil-like atomic ensemble along $\hat{x}$ direction and the wave vector $\mathbf{k}=k\hat{x}$. $|b_j\ra$ and $|c_j\ra$ are the excited and the ground states of the atom at position $x_j$. A uniform displacement of the atomic ensemble relative to the lab reference frame including all optical devices, $r_0$, attaches opposite phases to the two timed Dicke states \cite{Chen2010a}, $1/{\sqrt{2}}\left(e^{ikr_0}|b_\mathbf{k}\ra+e^{-ikr_0}|b_\mathbf{-k}\ra\right)$, which can be easily verified by replacing $x_j$ with $x_j+r_0$ in Eq. (\ref{sr}).
The phases can be retrieved by the interference pattern of the photon signal emitted in the two opposite directions, $\cos^2\left(kr_0\right)$. The whole process can be understood as follows. We first freeze the standing wave pattern composed by the entangled two modes $\pm\mathbf{k}$ in the atomic ensemble. Any displacement of the atomic ensemble moves the standing wave pattern stored in it simultaneously, which also correspondingly moves the interference pattern of the stored single photon after it is released. If we prepare the ensemble in an atomic N00N state with entangled $N$-fold collective excitations \cite{Thiel2007}, the interference pattern has $N$ times higher resolution \cite{Chen2010a}. However, the efficiency in preparing these repetitive N00N states drops exponentially with $N$ \cite{Hume2009, Chen2010a, Prevedel2009}.

In this Letter, we show that the ``momentum Schr\"odinger cat state''
\begin{equation}
|\text{KAT}(N)\ra=(-1)^N\frac{1}{\sqrt{2}}\left(|b_{-2N\mathbf{k}}\ra+i|a_{2N\mathbf{k}}\ra\right),
\label{cat}
\end{equation}
with large number $N$ can be prepared. Here ``a'' stands for another atomic level and $|a_{\mathbf{k}}\ra$ is defined in the same way as Eq. (\ref{sr}) by replacing $b$ with $a$. Instead of atomic N00N states with high atomic excitations, we only need single excitation superradiant states which can be prepared with high efficiency \cite{Lukin2003a}. We propose an experimental scheme based on ground states Raman transitions to avoid dephasing. 

\begin{figure}[t]
    \epsfig{figure=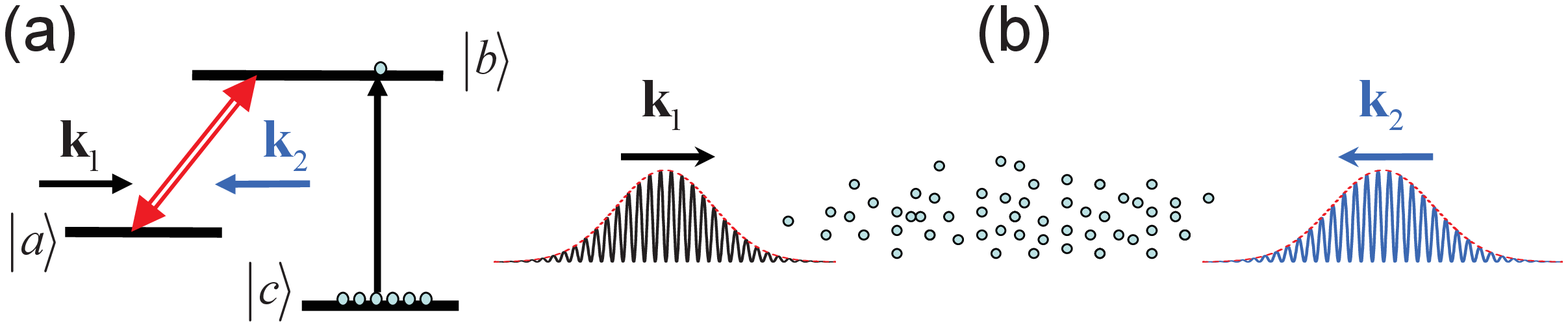, angle=0, width=0.48\textwidth}
    \epsfig{figure=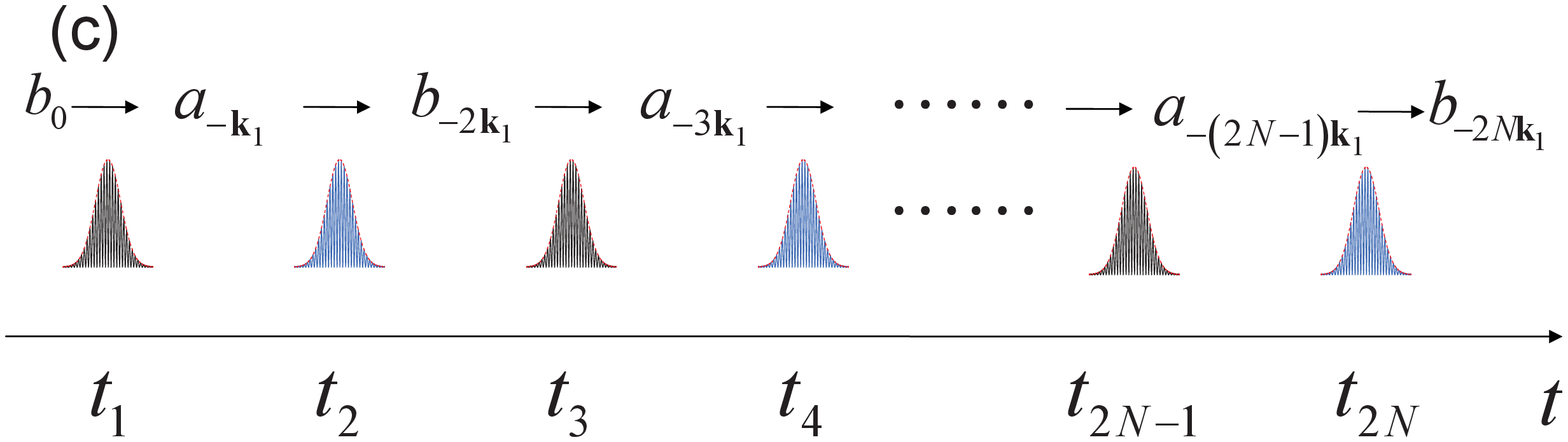, angle=0, width=0.48\textwidth}
    \epsfig{figure=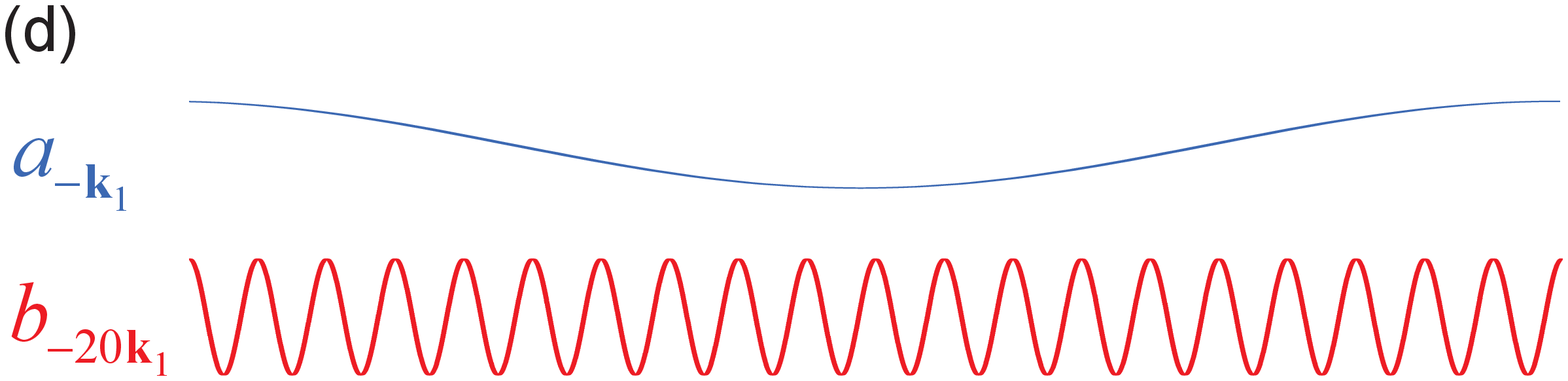, angle=0, width=0.48\textwidth}
\caption{(Color online) (a) Three-level configuration for photon momentum encoding. (b) Two counter-propagating modes collectively couple the transition between $|b\ra$ and $|a\ra$. (c) Timed Dicke state transport with $\pi$-pulses of $\mathbf{k}_1$ (black) and $\mathbf{k}_2$ (blue). They are applied alternatively to the atomic ensemble for $N$ times each to transfer the state from $|b_0\ra$ to $|b_{-2N\mathbf{k}_1}\ra$. (d) The schematic spatial phase oscillation of $|a_{-\mathbf{k}_1}\ra$ and $|b_{-20\mathbf{k}_1}\ra$.}
\label{pulse}
\end{figure}

\emph{Store multiple light momenta.}---To motivate the underlying physical mechanisms, we first show how to write multiple light momenta into an atomic ensemble. We use a three-level scheme as shown in Fig. \ref{pulse} (a). We first prepare the atomic ensemble in the superradiant state
$|b_0\ra=\frac{1}{\sqrt{N_a}}\sum_{j=1}^{N_a}\left|c_1,c_2,...,b_j,...c_{N_a}\right>$ 
by absorbing a single photon with momentum perpendicular to $\hat{x}$-axis. In that direction, the atomic medium is thin and uniform probability amplitude for each atom is easily achieved. Then we prepare two sequences of counter-propagating $\pi$-pulses in modes $\mathbf{k}_1=k_1\hat{x}$ and $\mathbf{k}_2=-k_1\hat{x}$ [Fig. \ref{pulse} (b)], and send them to the atomic ensemble alternatively, as shown in Fig. \ref{pulse} (c). These two modes couple the transition from $|b\ra$ to another state $|a\ra$, rather than the ground state $|c\ra$. The $\pi$-pulses are represented by the following unitary transform under the rotating-wave approximation
\begin{equation}
U_l
=\exp\left[i\frac{\pi}{2}\sum_{j=1}^{N_a}\left(e^{i k_l x_j+i\phi_l}\sigma_j^+ +e^{-i k_l x_j-i\phi_l}\sigma_j^-\right)\right],
\label{ul}
\end{equation}
where $l=1,2$ for forward and backward pulses. $\sigma_j^+=|b_j\ra\la a_j|$ and $\sigma_j^-=|a_j\ra\la b_j|$ are the raising and lowering operators for the $j$th atom. $\phi_l$ is the phase of field $l$. 

In the following, we set $\phi_1=\phi_2=0$. At time $t_1$, we apply $U_1$ to $|b_0\ra$, 
\begin{equation}
\begin{aligned}
U_1 |b_0\ra=i|a_{-\mathbf{k}_1}\ra.
\end{aligned}
\label{uni}
\end{equation}
The atomic ensemble transit to $|a_{-\mathbf{k}_1}\ra$ by collectively emitting a photon in mode $\mathbf{k}_1$, and therefore acquire momentum $-\hbar\mathbf{k}_1$ based on momentum conservation. In calculating Eq. (\ref{uni}) we should note that the terms $\sigma_j^{+,-}\sigma_{j^\prime\neq j}^{+,-}$ and the higher order ones in the expansion of $U_l$ applied to the single photon Dicke state lead to zero.

\begin{figure}[t]
    \epsfig{figure=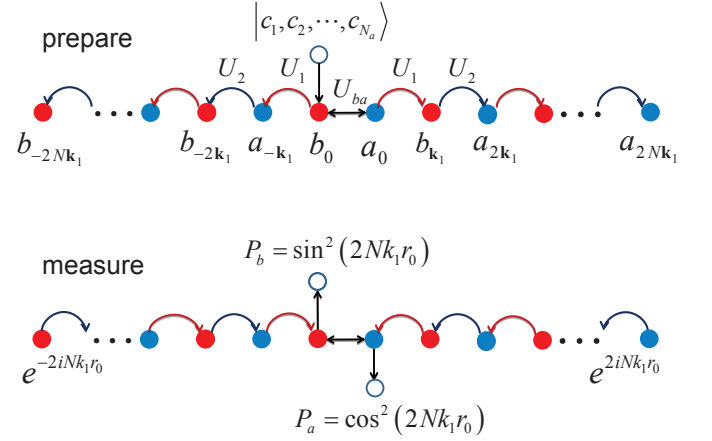, angle=0, width=0.5\textwidth}
\caption{(Color online) Pictorial scheme of the displacement metrology with ``momentum Schr\"odinger cat state'' based on Ramsey interferometry. The white circles represent the collective ground state $|c_1, c_2,..., c_{N_a}\ra$. The red and blue circles represent the superradiant sates $|b_{\mathbf{k}}\ra$ and $|a_{\mathbf{k}}\ra$. The single black arrow represent the single-photon Raman transition between the ground state and $|b_0\ra$ or $|a_0\ra$. The double black arrow represents the $\frac{\pi}{2}$-pulse $U_{ba}\left(\frac{\pi}{2}\right)$. The red and blue arrows represent the $\pi$-pulses $U_1$ and $U_2$ with the arrow's direction indicating the transition direction.}
\label{expschm}
\end{figure} 

At time $t_2$, we send the $\pi$-pulse of $\mathbf{k}_2$ and the state evolves to
\begin{equation}
U_2 i|a_{-\mathbf{k}_1}\ra=-|b_{-2\mathbf{k}_1}\ra.
\end{equation}
The atomic ensemble acquire another momentum $\hbar \mathbf{k}_2=-\hbar\mathbf{k}_1$ by collectively absorbing a photon from mode $\mathbf{k}_2$. The pulse pair $U_2U_1$ encode a total momentum $-2\mathbf{k}_1$ in the atomic ensemble. The above process can be repeated for another $N-1$ times and the final state becomes $\left(-1\right)^N |b_{-2N\mathbf{k}_1}\ra$ with a large effective momentum $-2N\hbar\mathbf{k}_1$. Instead of using recoil momenta like in the atom interferometry \cite{Chiow2011}, the large momentum stored in the atomic ensemble is transferred to rapid oscillations of the phase correlation of timed Dicke states, as shown in Fig. \ref{pulse} (d). The enhanced oscillation allows improved precision for measuring the displacement. It is as if the ruler has a finer graduation.

\emph{Superradiant metrology.}---By combining the above mechanism and the technique of Ramsey interferometry, we can measure a displacement to the Heisenberg limit. The whole scheme is sketched in Fig. \ref{expschm}. We first prepare a superposition state of $|b_0\ra$ and $|a_0\ra$. Then the $\pi$-pulses drive these two states in two opposite directions in momentum space to obtain a ``momentum Schr\"odinger cat state''. 

\begin{figure}[t]
    \epsfig{figure=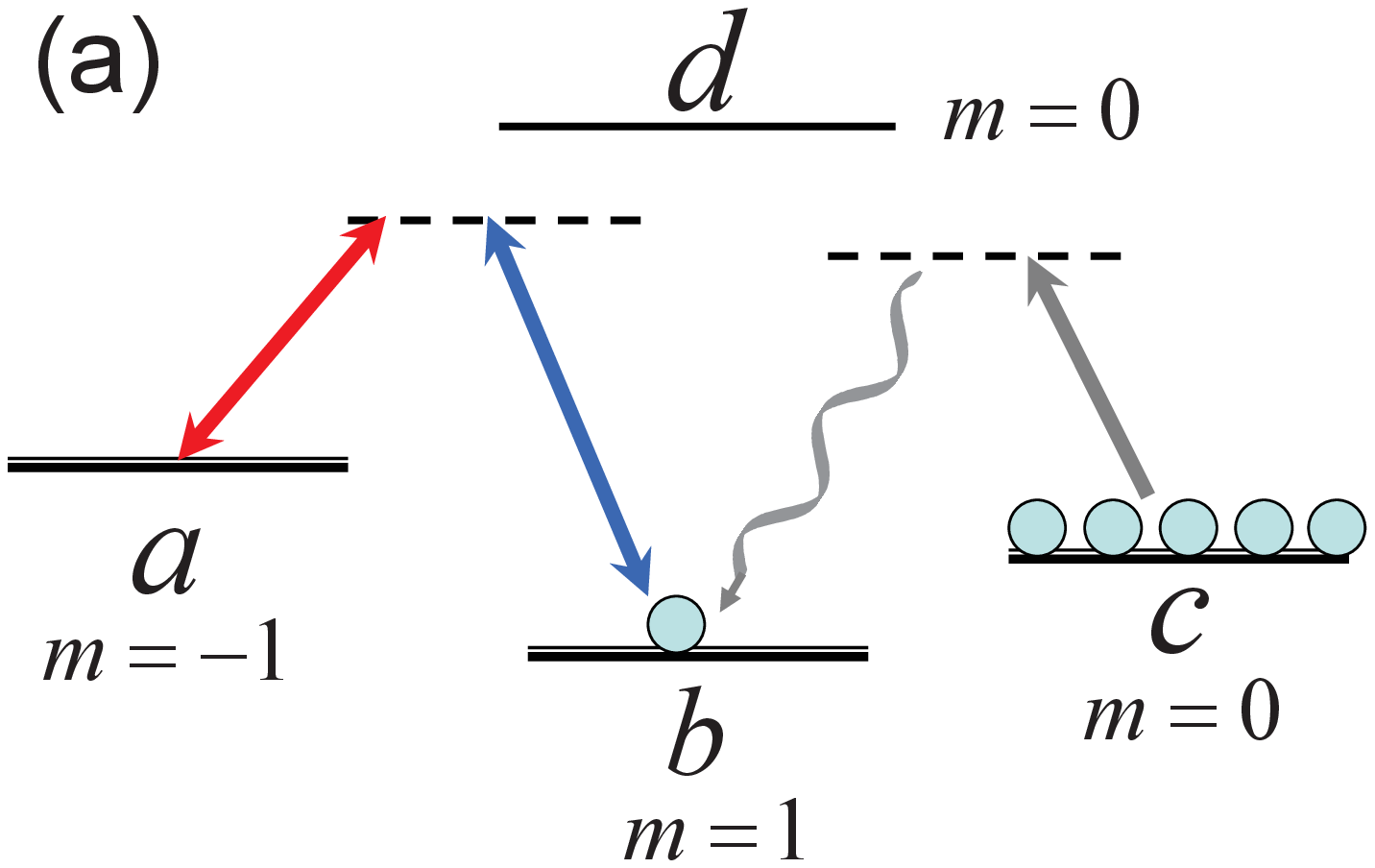, angle=0, width=0.23\textwidth}
    \epsfig{figure=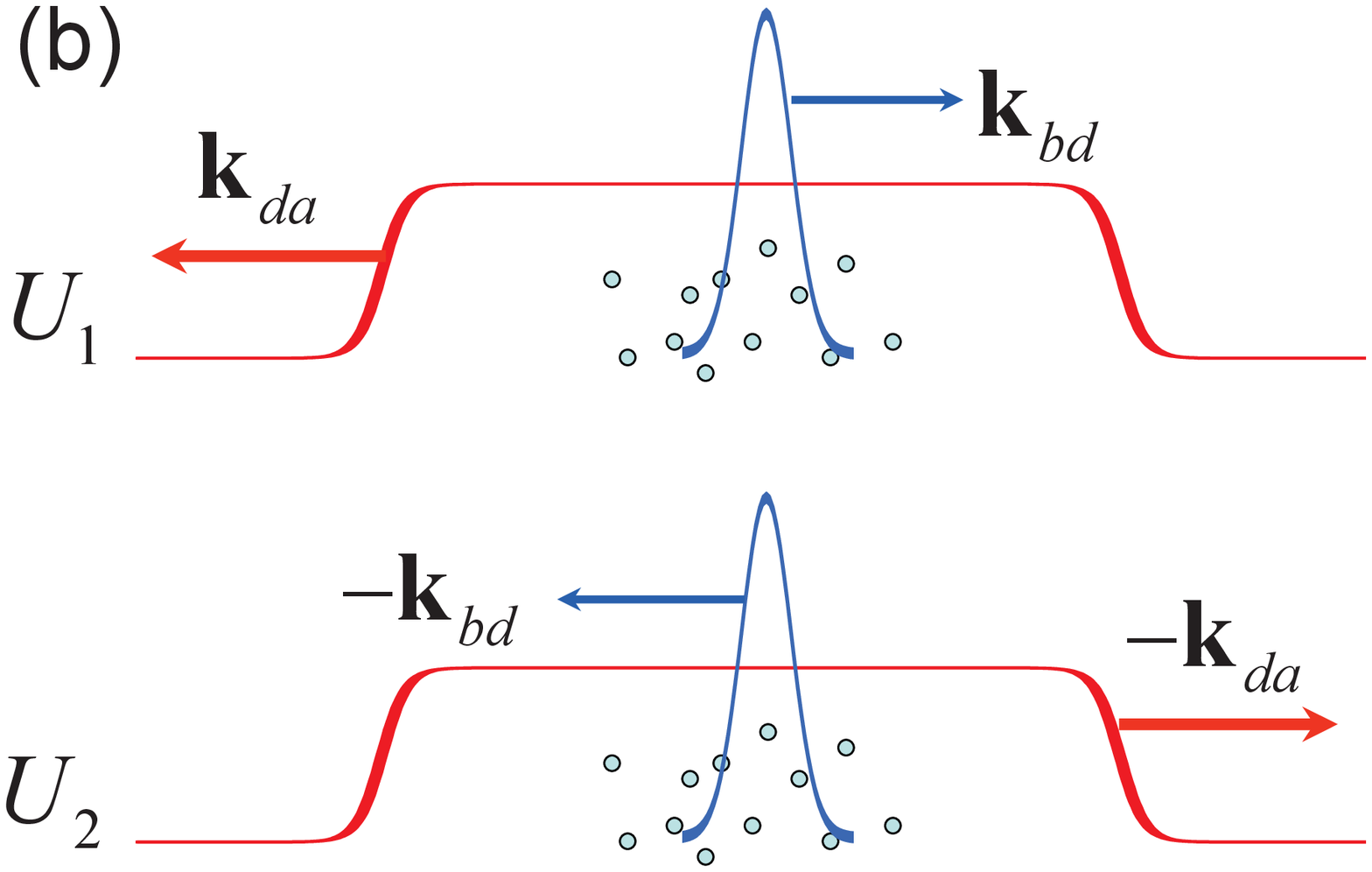, angle=0, width=0.24\textwidth}
\caption{(Color online) Superradiant metrology with Raman transitions. (a) Atomic levels of Raman transitions. $|a\ra$, $|b\ra$ and $|c\ra$ are now the three Zeeman ground state sublevels. The atomic ensemble are initially prepared in state $|c\ra$. A single photon Raman transition (green arrows) prepares the atomic ensemble in a timed Dicke state of $|b\ra$. Then the Raman pulses drive the transition between $|b\ra$ and $|a\ra$. (b) Raman $\pi$-pulses $U_1$ and $U_2$. They are composed by a long pulse driving $d\leftrightarrow a$ transition and a short pulse driving $b \leftrightarrow d$ transition. The short pulse goes through the atomic ensemble when the atoms are uniformly covered by the long pulse. }
\label{raman}
\end{figure}

We show the explicit procedure based on a three-level Raman configuration which has been proved to have decoherence time as long as 1 minute \cite{Dudin2013, Heinze2013}. The atom has three degenerate ground states which can be lifted by a Zeeman magnetic field along $\hat{x}$, as shown in Fig. \ref{raman} (a). We first pump all the atoms to state $|c\ra$. An off-resonant coherent field with $\hat{x}$-polarization induces a Raman transition via intermediate state $|d\ra$ to prepare the atomic ensemble in the state $|b_0\ra$, accompanied by the emission of a right circular polarized Stokes photon \cite{Duan2001, Chen2010a, Chou2004}, whose frequency and polarization can be detected as signatures of a successful preparation of the state $|b_0\ra$. A Raman $\frac{\pi}{2}$-pulse \cite{Braje2014} which couples the transition from $|b\ra$ to $|a\ra$ transforms $|b_0\ra$ to
\begin{eqnarray}
\begin{aligned}
U_{ba}\left(\frac{\pi}{2}\right)|b_0\ra=\frac{1}{\sqrt{2}}\left(|b_0\ra+i|a_0\ra\right),
\end{aligned}
\label{ini}
\end{eqnarray}
where $U_{ba}\left(\frac{\pi}{2}\right)=\sum_{j=1}^{N_a}\left(\cos\frac{\pi}{4}I_j+i\sin\frac{\pi}{4}\sigma_j^x\right)$. Now we introduce Raman transitions with combined wave vectors $\mathbf{k}_1=\mathbf{k}_{bd}-\mathbf{k}_{da}$ and $\mathbf{k}_2=-\mathbf{k}_1$, as shown in Fig. {\ref{raman}} (b). Here $\mathbf{k}_{bd}$ and $\mathbf{k}_{da}$ are the wave vectors of the two circularly polarized modes which couple $|b\ra\leftrightarrow|d\ra$ and $|d\ra\leftrightarrow|a\ra$ transitions respectively. We adopt a counter-propagating configuration along $\hat{x}$-axis to achieve a maximum combined wave vector \cite{Longdell2005}. 

The $\pi$-pulses of these Raman transitions can still be represented by Eq. (\ref{ul}). We apply $U_1$ to the state in Eq. (\ref{ini}) and we obtain
$1/{\sqrt{2}}\left(i|a_{-\mathbf{k}_1}\ra-|b_{\mathbf{k}_1}\ra\right)$. 
We then apply $U_2$ to the above sate and we obtain
$-1/{\sqrt{2}}\left(|b_{-2\mathbf{k}_1}\ra+i|a_{2\mathbf{k}_1}\ra\right)$.
Repeating the above operations another $N-1$ times, we get the ``momentum Schr\"odinger cat state'',
\begin{equation}
|\text{KAT}(N)\ra=\left(-1\right)^N\frac{1}{\sqrt{2}}\left(|b_{-2N\mathbf{k}_1}\ra+i|a_{2N\mathbf{k}_1}\ra\right).
\label{kat}
\end{equation}
Now we move the atomic ensemble collectively by a distance $r_0$ with a uniform optical force or gravity, etc. \cite{Chen2010a}, so that the old position $x_j= x^\prime_j+ r_0$ where $x^\prime_j$ is the new position. $|\text{KAT}(N)\ra$ is expressed in Eq. (\ref{fin}) with this displacement. We must redefine the timed Dicke states with the new positions of the atoms in the lab frame. Replacing $x_j$ with $x^\prime_j+r_0$, Eq. (\ref{kat}) becomes
\begin{equation}
\left(-1\right)^N\frac{1}{\sqrt{2}}\left(e^{-i2Nk_1r_0}|b_{-2N\mathbf{k}_1}\ra+ie^{i2Nk_1r_0}|a_{2N\mathbf{k}_1}\ra\right),
\label{fin}
\end{equation}
where the two timed Dicke states are redefined with $x^\prime_j$ and have a relative phase $4Nk_1r_0$. The center-of-mass Hamiltonian only brings trivial global phases. We neglect the dynamic phase difference, which can be easily compensated in experiments or data analysis. To retrieve the phase, we apply the inverse $\pi$-pulse sequences $\left(U_1U_2\right)^N$ to the state in Eq.(\ref{fin}). Because the unitary transform is reversible, we will finally get
\begin{equation}
|\Psi\ra=\frac{1}{\sqrt{2}}\left(e^{-i2Nk_1r_0}|b_0\ra+ie^{i2Nk_1r_0}|a_0\ra\right).
\label{fin1}
\end{equation}
A Raman $\frac{\pi}{2}$-pulse $U_{ba}(\frac{\pi}{2})$ transforms the above state to
\begin{equation}
-i\sin\left(2Nk_1r_0\right)|b_0\ra+i\cos\left(2Nk_1r_0\right)|a_0\ra.
\end{equation}
The probability of the state $|b_0\ra$, $P_b=\sin^2\left(2Nk_1r_0\right)$ can be obtained by observing the retrieved photon via the forward Raman transition $|b\ra\rightarrow|d\ra\rightarrow|c\ra$ after a pumping pulse coupling $|b\ra$ to $|d\ra$ is applied along $\hat{x}$. Here the Raman transition $|b\ra\rightarrow|d\ra\rightarrow|c\ra$ will happen rather than $|b\ra\rightarrow|d\ra\rightarrow|a\ra$ due to a superradiant enhancement of the vacuum interaction for the former one \cite{Duan2001}. 
The probability of $|a_0\ra$, $P_a=\cos^2\left(2Nk_1r_0\right)$ can be simultaneously measured in a different direction. The population difference
\begin{equation}
P=P_b-P_a=-\cos\left(4Nk_1r_0\right)
\label{sgn}
\end{equation}
is the signal from which the displacement $r_0$ can be measured. The noise is $\Delta P=\left|\sin\left(4Nk_1r_0\right)\right|$ and the phase sensitivity
\begin{equation}
\Delta (k_1r_0)=\frac{\Delta P}{\left|\partial P/\partial (k_1r_0)\right|}=\frac{1}{4N}
\label{sens}
\end{equation}
scales at the Heisenberg limit.

\emph{Discussion.}---It has been argued that the sensitivity in Eq. (\ref{sens}) only shows the accuracy in determining the last digits of $k_1r_0$ due to the periodicity of the signal $P$. However, we can take some iterative procedure to determine all the digits without destroying the $1/N$ scaling \cite{Giovannetti2006, Xiang2011}. Especially, to verify existing theories like gravitational wave, the phase change from the theoretical prediction is small enough to be within half period of the signal $P$. The improvement of the sensitivity can be seen by expanding $P_b$ near $r_0=0$, $P_b\approx 4N^2k_1^2r_0^2$. The probability of detection is enhanced by $N^2$. As in most interferometry experiments, $r_0$ can be changed continuously and the relevant physical quantity like the gravitational constant can be measured from the interference pattern rather than a single point.

\begin{figure}[t]
    \epsfig{figure=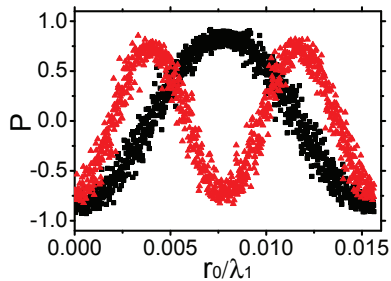, angle=0, width=0.3\textwidth}
\caption{(Color online) Numerical simulation of $P$ for Gaussian noisy $\pi$-pulses with area variation $\Delta S=0.1$ and phase variation $\Delta \phi=0.01$. $N=16$ (black squares) and $32$ (red triangles).}
\label{intf}
\end{figure}

The imperfection of the $\pi$-pulses in amplitude and in phase due to environmental noises, such as oscillations and rotations of optical devices, can reduce the phase sensitivity. We suppose the area $S$ and the phase $\phi$ of the $\pi$-pulses have Gaussian distribution with variations $\Delta S,\Delta \phi\ll 1/\sqrt{N}$. The phase sensitivity is then $\Delta \left(k_1r_0\right)=\left[4N\left(1-N\Delta S^2/2\right)\right]^{-1}+\Delta\phi/\sqrt{4N}$. We simulate the interference patterns in Fig. \ref{intf} for $\Delta S=0.1$ and $\Delta \phi=0.01$. The super-resolving metrology is demonstrated by the reduction of the oscillation period of the interference pattern to $\lambda_1/4N$ where $\lambda_1=2\pi/k_1$ is the effective wavelength. Although $\Delta S$ reduces the visibility and $\Delta \phi$ blurs the interference pattern, the phase sensitivities for $N=16$ and 32 are still enhanced to 1/55 and $1/98$, marginally lower than the Heisenberg limit $1/64$ and $1/128$ whereas much higher than the shot-noise limit $1/8$ and $1/11$.

The pure dephasing between the ground states due to environmental noise fields does not make the ``momentum Schr\"odinger cat state'' more fragile as the number $N$ increases. Although there are $2N$ light momenta encoded in the media, the ensemble only contain a single excitation whose dephasing is independent of $N$. The relative motion between the atoms will reduce the visibility by a factor $e^{-\epsilon^2 N^2}$ as we will discuss later. Therefore, to propose an experimental implementation, a solid system where the relative distance between atoms are fixed is preferred, especially the earth-ion-doped crystal, such as Pr$^{3+}$:Y$_2$SiO$_5$ whose quantum memory storage time reaches to 1 minute \cite{Longdell2005, Heinze2013}. The three states $|a\ra$, $|b\ra$ and $|c\ra$ can be chosen from the ground state $^3H_4$. The intermediate state $|d\ra$ is a sublevel of the excited state $^1D_2$. The Zeeman splitting can be $\sim$ MHz. The control pulses of $U_{1,2}$ can have duration of 10$\mu$s to avoid transition to unwanted levels. The prepare and read stage of the measurement can cost time 10ms for $N=10$. Then for the transition wavelength $\sim$600nm, the resolution is $7.5$nm. The crystal can free fall 100nm within a millisecond in gravity field. Therefore the whole measurement can be completed well within the decoherence time of 1 minute.

For a proof of principle verification of the mechanism, the ultra cold atoms can also work. The obstacle is that the thermal motions of the atoms will randomize the relative phase between the atoms, and reduce the visibility $P=-e^{-2(Nk_1v_m\tau)^2}\cos\left(4Nk_1r_0\right)$ where $v_m=\sqrt{2k_B T/m}$ is the most probable velocity and $\tau$ is the overall time cost by the measurement. The phase sensitivity is thus 
$\Delta\left(k_1r_0\right)=e^{\epsilon^2 N^2}/4N$,
where $\epsilon=\sqrt{2}k_1v_m\tau$, roughly the number of wavelengths the atoms travelled. We use a copropagating configuration in the Raman pulses to achieve a small $k_1$ and consequently a small $\epsilon$. Take $^{87}$Rb as an example \cite{Chen2010a, Dudin2013}, the three levels are chosen to be the three Zeeman sublevels of 5$^2$S$_{1/2}$, $F=1$. The intermediate state is 5$^2$P$_{1/2}$, $F=2$, $m_F=0$. If we use nanosecond Raman transition pulses with GHz positive detuning, the $\pi$-pulses require a moderate average power 10$\text{W}/\text{cm}^2$. At temperature $T\sim\mu\text{K}$, $v_m\sim 1\text{cm}/\text{s}$. We suppose the whole measurement costs time $\tau=100\mu$s during which a displacement of  10$\mu$m can be achieved by optical force \cite{Chen2010a}, whereas the random displacement is $v_m\tau\sim 1\mu$m. If the effective wavelength in the Raman transition is $\lambda_1\sim 200\mu$m, for resolution of $5\mu$m, we need $N=10$. The interference pattern becomes $P\approx -0.82\cos\left(40k_1r_0\right)$. The phase sensitivity $1/32$ still exceeds the shot-noise limit $1/\sqrt{40}\sim 1/6$.

The conventional Heisenberg limit metrology by entangling $N$ atoms or photons is difficult to be realized when $N$ is large. Recent developments in this field achieved 5-photon N00N states \cite{Afek2010b}, 8-photon GHZ states \cite{Huang2011} and 14-ion GHZ states \cite{Monz2011}. The fast decoherence of multi-particle entangled states limits their realization of large $N$. Our protocol circumvents this obstacle by consuming the quantum resources without directly entangling them. The physical quantity that we use is the light momentum. We consume one photon from each light pulse and store its momentum as a frozen spin wave of atoms. This storage has been proved to be very robust \cite{Lukin2003}. Our single excitation scheme has the following advantages. First, its decoherence rate is independent of $N$, the number of the quantum resources we consumed. Second, we avoid the difficult collective detection of $N$ particles, which is usually required for N00N states and GHZ states. Third, we do not need multiple passes of the probe such as in the single photon entanglement-free scheme \cite{Higgins2007}.

Atoms are not consumed in the measurement. The number of atoms $N_a$ has no relation with the $1/N$ scaling. We need $N_a$ to be large (usually in the order of $10^6$ for cold atoms) and distributed in a pencil-like region much longer than $\lambda_1$ for good directionality of the signal photon. We do not have atom loss if we use crystals. For cold atoms, if the remaining atoms are $\eta N_a$, the signal is reduced to $\eta P$, which has no influence on the scaling.

In conclusion, we propose a Heisenberg limit metrology with an atomic ensemble in a ``momentum Schr\"odinger cat state'' prepared by encoding multiple photon momenta into the phase correlations of timed Dicke states. We analysed the feasibility of proof-of-principle experiments in Pr$^{3+}$:Y$_2$SiO$_5$ crystals and in ultra cold $^{87}$Rb atoms. Since light momenta transfer to atoms already exceeds 100 with current technology \cite{Chiow2011}, our protocol is promising to improve the scalability of Heisenberg limit metrology by one order.

We thank W. P. Schleich, R. Glauber, S.Y. Zhu, J. Evers, C. O'Brien, M. Kim, L. Yuan, K. Wang, and H. Cai for helpful discussions. We gratefully acknowledge the support of the National Science Foundation Grants No. PHY-1241032 (INSPIRE CREATIV) and PHY-1068554 and the Robert A. Welch Foundation (Grant No. A-1261).

\bibliographystyle{nature}
\bibliography{hopbg}

\end{document}